\documentclass[twocolumn,aps,pre]{revtex4}
\usepackage{graphicx,color}
\begin{document}
\title{\bf{Length dependent dynamics of microtubules}}
\author{Vandana Yadav and  Sutapa Mukherji}
\affiliation{Department of Physics, Indian Institute of Technology,
Kanpur-208 016}
\date{\today}
\begin{abstract}
 
Certain regulatory proteins influence the polymerization dynamics of 
microtubules by inducing catastrophe with a rate that depends on the 
microtubule length. 
Using a discrete formulation, 
here  we show that, for catastrophe rate proportional to the 
microtubule length,  the steady-state  
probability distributions of length   decay   much faster with length 
than an exponential decay as seen in the absence of these proteins. 
\end{abstract}
\maketitle
\section{Introduction}
 Microtubules are   important 
components of the cytoskeleton of the cell \cite{lodish}. These 
are hollow microscopic tubes  formed by 
$\alpha$,  $\beta$-tubulin heterodimers. These dimers are joined end-to-end
to form protofilaments with alternating $\alpha$, $\beta$ subunits. 
The wall of the microtubule  
has a staggered arrangement of typically  $13$ 
such protofilaments.  Experimental 
observations of the polymerization dynamics of microtubules reveal that 
microtubules alternate between persistent polymerizing and depolymerizing 
phases. This unusual feature is known as the "dynamic instability" 
\cite{mitchison}. 
The transition from a polymerizing to a depolymerizing phase is known 
as "catastrophe" and the same from a depolymerizing phase to a 
polymerizing phase is known as "rescue". A popular idea regarding the 
 cause of the  dynamic instability is that it 
occurs as a result of two competing processes, the addition of 
GTP bound tubulins at the growing tip and the hydrolysis of the GTP units 
of the tubulins present on the microtubule \cite{mitchison,pantaloni,fly}. 
The GTP bound tubulins provide a stabilizing cap to the 
growing end and if this cap is lost due to  rapid hydrolysis of the GTP 
units, the microtubule enters into a depolymerizing phase. 

Microtubules are responsible  for various intracellular 
organisation,  exerting pushing or pulling forces etc. 
It is, therefore, important that the 
polymerization  and depolymerization of microtubules are 
appropriately 
regulated  so that the length distribution of microtubules satisfies 
the requirement of the cell. It is found that the  microtubule dynamics is 
linked to a large number of regulatory mechanisms of microtubule 
associated proteins \cite{lodish}.   {\it In vivo} experiments show that 
 a variety of   proteins  known as polymerases or 
depolymerases  stabilize or destabilize 
the microtubules by 
reducing or enhancing the catastrophe rate \cite{howard} respectively.   
 Although the mechanism as exactly how the microtubule associated
proteins influence the dynamics is not known, there are experimental 
evidences of length dependent regulation of microtubule dynamics under 
different cellular contexts \cite{dogterom1,gardner,dogterom2,varga}. 
While in certain cases, there are speculations about the microscopic 
origin of such regulations \cite{dogterom1}, recent 
{\it in vivo} studies on two different systems find  depolymerases
belonging to kinesin-5 \cite{gardner} and kinesin-8 
\cite{dogterom2} family of proteins 
responsible for such length-dependent 
regulation.  In particular,  
 live-cell imaging of interphase fission yeast   
cells \cite{dogterom2} suggests that kinesin-8 regulation 
leads to an increase of  the catastrophe 
frequency with the microtubule length. 
Since, in this case,  the depolymerization is length dependent, 
 cells may selectively control the 
(de)polymerization dynamics of different 
populations of microtubules.  As a consequence, 
the activity of such depolymerases is  important for  
formation of ordered  microtubule-based structures during cell 
division.
Recent numerical simulations suggest that such length 
dependent regulation of polymerization dynamics is crucial 
also for microtubule related processes during cell division 
\cite{odde1}.

A possible mechanism  through which kinesin-8 might introduce 
a length dependence in the catastrophe frequency has recently been 
proposed in \cite{nedelec}. 
The authors of \cite{nedelec} simulate the regulatory mechanism 
based
on the following postulates. 
The processive depolymerases, after binding,  walk towards the 
plus end of the microtubule and accumulate near its tip.  
Since these molecules can bind anywhere along the 
length with equal 
probability, and unbind rarely, it is likely that longer microtubules
have larger density of the proteins at the tip than the shorter ones.  
Once the  motor reaches the tip, it  encounters  a 'push' from the 
subsequent motors. A 'pushed' motor at the tip finally breaks off 
taking  the terminal tubulin along with. 
This  gives rise to catastrophe due to 
the  shortening of the GTP cap. 
These studies indicate  that the  catastrophe frequency 
increases linearly with the length of the 
the microtubule, as observed experimentally from live-cell 
imaging of fission yeast cells \cite{dogterom2}.

In the context of such dynamic instability, the central properties 
of interest are the length distributions of a set of 
microtubules \cite{hill,leibler93}. 
Apart from being useful for obtaining various 
statistical averages as a function of the rate parameters, these 
distributions lead to predictions of various quantities 
such as microtubule's mean life time, elongation time \cite{rubin},
 mean chromosome search time \cite{bindu} etc.
  that can be tested experimentally.  
 In this work, by extending the two state model proposed earlier by 
Hill \cite{hill}, we study how the length dependent catastrophe rate 
affects the  length distribution of 
microtubules.
Motivated by their own observation, authors of \cite{dogterom2} 
have proposed a length-dependent  two-state model in which 
the catastrophe rate depends linearly on the microtubule 
length \cite{dogterom3}.  Following a continuum approach, 
they obtain microtubule length distributions and use these distributions  
further to obtain  positional 
information regarding the preferential cite for the future 
growth of the cell. In view of the relevance of the 
distributions, we are motivated to adopt a  discrete formulation 
to obtain exact expressions of various 
length distribution functions. The length distribution obtained in 
\cite{dogterom3} has a Gaussian form and our results 
agree with those of \cite{dogterom3} 
under a special condition satisfied by the growth and catastrophe rates 
and the microtubule length.

The discrete description, unlike the continuum approach involving two 
partial differential equations, requires two infinite sets   
of coupled first order dynamical equations. Since further 
consideration of the internal protofilament structure of the 
microtubule would lead 
to more complexity, we have considered a microtubule to be a 
polymer of a single protofilament 
 as a first approximation. Given the fact that 
the longitudinal bonds within a protofilament  are strong
in comparison with the lateral interaction \cite{odde}, 
it is possible to extend the model 
by taking neighboring protofilaments into account. However, it may be 
worth  noting that considering a "gating rescue" effect from the 
neighboring protofilament as done for the composite model of 
\cite{nedelec} leads to a behavior qualitatively similar to the 
single protofilament case. 
Other models   of polymerization dynamics 
 with length dependent attachment and detachment rates exist 
\cite{gonzalez}, but these models 
do not take into account the persistent nature of growth and shrinkage.   
In view of this, the present  analysis,   
allows a  direct comparison with the 
results of \cite{hill} in the steady-state.

The evolution equations 
are written for $p_n^+$ and $p_n^-$ which are probability densities 
of having microtubules of length $n$ in a growing or a shrinking phase 
respectively.   
The evolution equations for these  probability densities are 
\begin{eqnarray}
&&\frac{dp_n^+}{dt}=r_a  p_{n-1}^++r_{-+}p^-_n-(r_{+-}n+r_a )p_n^+,\ 
\ \ n\ge 2 \label{polymer+},\nonumber\\
\label{polymer+}\\
&&\frac{dp_n^-}{dt}=r_d p_{n+1}^-+r_{+-}np_n^+-(r_d+r_{-+})p_n^
-,\ n\ge 1 \label{polymer-}.
\end{eqnarray} 
Here, $r_a$, $r_d$, $r_{+-}$ and $r_{-+}$  denote 
the tubulin attachment 
rate, tubulin detachment rate, catastrophe rate and the rescue rate 
respectively. As seen in these equations, the catastrophe 
rate, $r_{+-}n$, for transition from 
$p_n^+$ to $p_n^-$ is proportional to the microtubule length, $n$.

\section{Length independent catastrophe rate: correspondence with 
previous results}

Here we consider the case where the catastrophe rate is independent of 
the length of the microtubule  and find correspondences with  
the  results  of reference {\cite{hill}}.
The steady-state recurrence relations satisfied by the probability 
densities are 
\begin{eqnarray}
r_a p_{n-1}^++r_{-+} p_n^--(r_{+-}+r_a) p_n^+=0 \ \  {\rm for}\ \  n\ge 2\\
r_d p_{n+1}^-+r_{+-} p_n^+-(r_d+r_{-+}) p_n^-=0 \ \ {\rm for}\ \  n\ge 1.
\end{eqnarray}
Initially, we assume that $p_1^+$ remains constant at a given value. 
We use the generating function method to find the 
dependence of the  probability densities on the microtubule length. 
The generating functions defined as 
\begin{eqnarray}
G(x)=\sum_{n=0}^\infty x^n p_{n+2}^+,\ \ 
 H(x)=\sum_{n=0}^\infty x^n p_{n+2}^-\label{gh}
\end{eqnarray}
satisfy the following equations 
\begin{eqnarray}
&& [r_a x-(r_{+-}+r_a)]G(x)+r_{-+} H(x)=-r_a p_1^+ \\
&& [r_d-(r_d+r_{-+})x]H(x)+r_{+-} x G(x)=(r_d+r_{-+}) p_1^-\nonumber\\
&& -r_{+-} p_1^+.
\end{eqnarray}
Writing these  equations in the matrix form as    
\begin{eqnarray}
{\cal A} {\cal G}={\cal B},\label{matrixsoln}
\end{eqnarray}
where 
\[{\cal G}= \left( \begin{array}{c}
G(x)\\
H(x)  \end{array} \right),\] \ \ \ 
\[{\cal A}= \left( \begin{array}{cc}
r_a x-(r_{+-}+r_a)  & r_{-+}  \\
r_{+-} x  & r_d-(r_d+r_{-+}) x  \end{array} \right)\]  \ \ \ {\rm and}\\
\[{\cal B}= \left( \begin{array}{c}
-r_a p_1^+ \\
(r_d+r_{-+})p_1^--r_{+-} p_1^+\end{array} \right),\]
we find  the solutions for  $G(x)$ and $H(x)$ as 
\begin{eqnarray}
&& G(x) =\frac{1}{ \mid {\cal A}\mid} [-\{r_d-(r_d+r_{-+})x\} r_a p_1^++
r_{+-} r_{-+} p_1^+\nonumber\\&& -r_{-+}(r_d+r_{-+}) p_1^-], \label{geqn}\\
&& H(x)=\frac{1}{\mid {\cal A}\mid}[r_{+-} r_a p_1^+ x+
\{r_a x-(r_a+r_{+-})\}\times\nonumber\\ 
&& \{(r_d+r_{-+})p_1^--r_{+-} p_1^+\}],\label{heqn}
\end{eqnarray}
where $\mid {\cal A}\mid=(1-x)[r_a(r_d+r_{-+})x-r_d(r_a+r_{+-})]$ is the 
determinant of  matrix ${\cal A}$. Clearly, matrix ${\cal A}$ is singular 
at $x=1$ and this poses a  question on the basic assumption of the 
steady-state. However, this singularity is a removable one if one considers 
the  time evolution of the probability density, $p_1^+$,  
and the probability density of nucleating 
sites $p_0$. With the following evolution equations \cite{hill}, 
\begin{eqnarray}
&&\frac{dp_1^+}{dt}=r_a p_0+r_{-+}p_1^--(r_a+r_{+-}) p_1^+\\
&&\frac{dp_0}{dt}=-r_a p_0+r_d p_1^-,
\end{eqnarray}  
one obtains the  steady-state conditions $p_1^-=\frac{r_a}{r_d} p_0$ and 
 $p_1^+=D_0 p_0$, with
$D_0=\frac{r_a(r_{-+}+r_d)}{r_d(r_{+-}+r_a)}$.
It is straightforward to see that the singularities 
in (\ref{geqn}) and (\ref{heqn})  disappear
once the ratio  $p_1^+/p_1^-=\frac{r_{-+}+r_d}{r_{+-}+r_a}$
 is    substituted in (\ref{matrixsoln}). 
Subsequent algebra leads to the 
steady-state probability densities 
\begin{eqnarray}
p_n^+=D_0^n p_0, \ \ {\rm and} \ \  
p_n^-=\frac{r_a}{r_d} D_0^{n-1} p_0 \label{ratioindependent}
\end{eqnarray}
as given in \cite{hill}.  It is important to note that 
the ratio  $p_n^+/p_n^-$ as obtained here does  not  match 
with the results of Appendix A of 
\cite{dogterom3}. (Also see the discussion below.) 
  The steady-state 
probability density $p_n^+$ decays exponentially with $n$  as 
\begin{eqnarray}
p_n^+=p_0 \ \exp[-\alpha n], \ \  {\rm with}\ \  
\alpha= -\ln{D_0}, \label{pnpexpo}
\end{eqnarray}
 if    $\alpha$ is positive.
In general, with constant  $p_1^+$,  the singularity at $x=1$ 
is  removable provided  $p_1^-$ evolves with 
time such that the  above ratio of $p_1^+/p_1^-$ is satisfied. 
This ratio is approached as 
$t\rightarrow \infty$  with a correction disappearing 
exponentially with time.

\section{Length dependent catastrophe rate}
\begin{figure}[htbp]
  \begin{center}
   \narrowtext
\includegraphics[height=2.5 in,clip, angle=0]{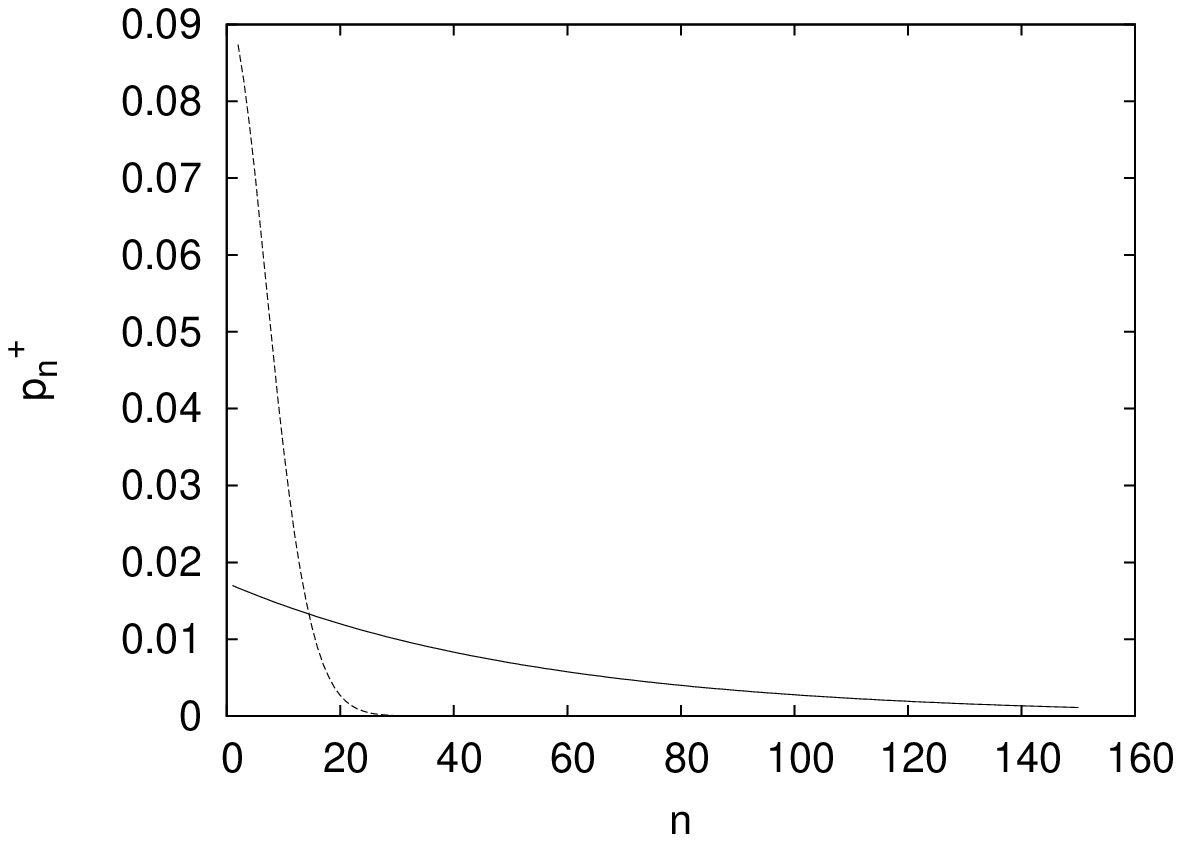}
    \caption{Steady-state  probability 
densities for growing microtubules are plotted with length.
 Solid and dashed lines 
correspond to  length independent and length dependent catastrophe rates 
respectively. Parameter values used for the plot are $r_a=10 \ s^{-1}$, 
$r_d=40\ s^{-1}$, $r_{-+}=0.02\ s^{-1}$ and 
$r_{+-}=0.19\ s^{-1}$.}\label{fig:valuep1p}
\includegraphics[height=2.5 in,clip,angle=0]{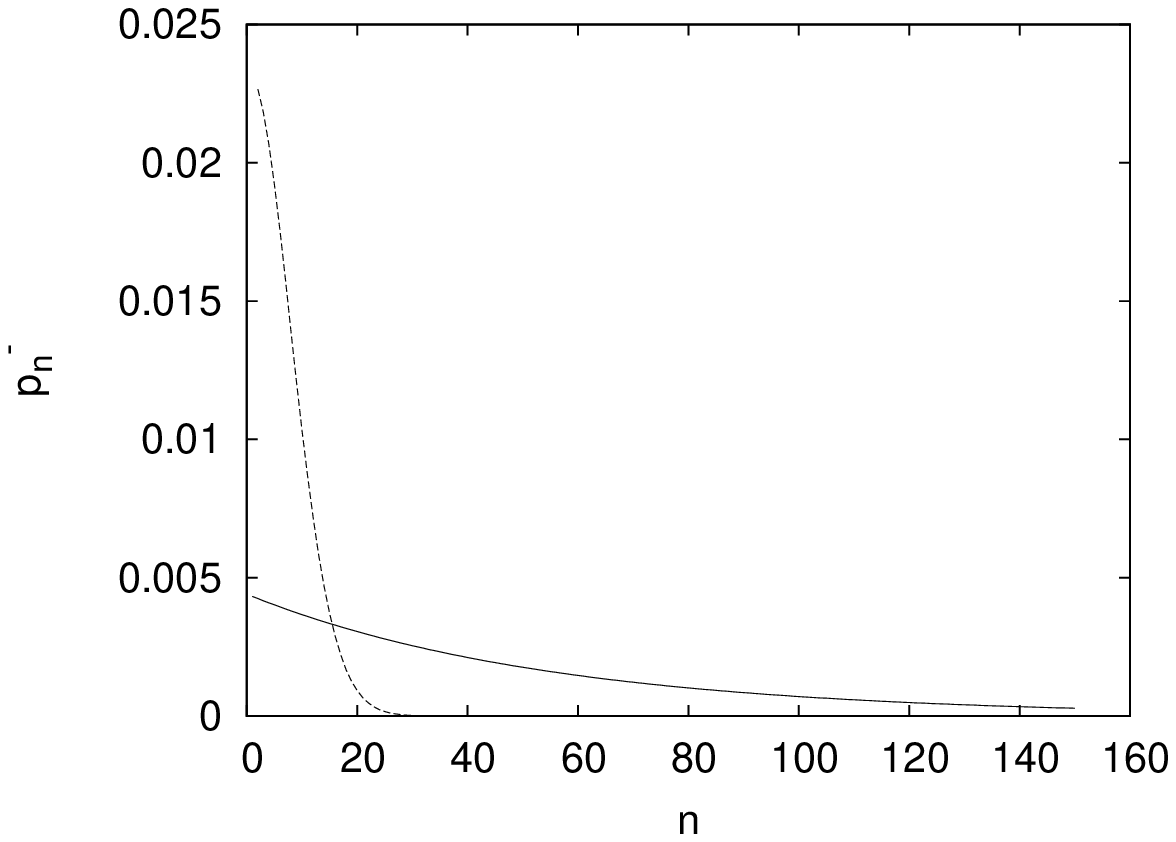}
\caption{Steady-state probability densities for shrinking microtubules 
are plotted  with length. Solid and dashed lines have the same meaning  as 
in Fig \ref{fig:valuep1p}. Parameter values are same as those of  
Fig. \ref{fig:valuep1p}.}
\label{fig:valuep1m}
  \end{center}
\end{figure}

\begin{figure}[htbp]
  \begin{center}
   \narrowtext
\includegraphics[height=2.5 in,clip, angle=0]{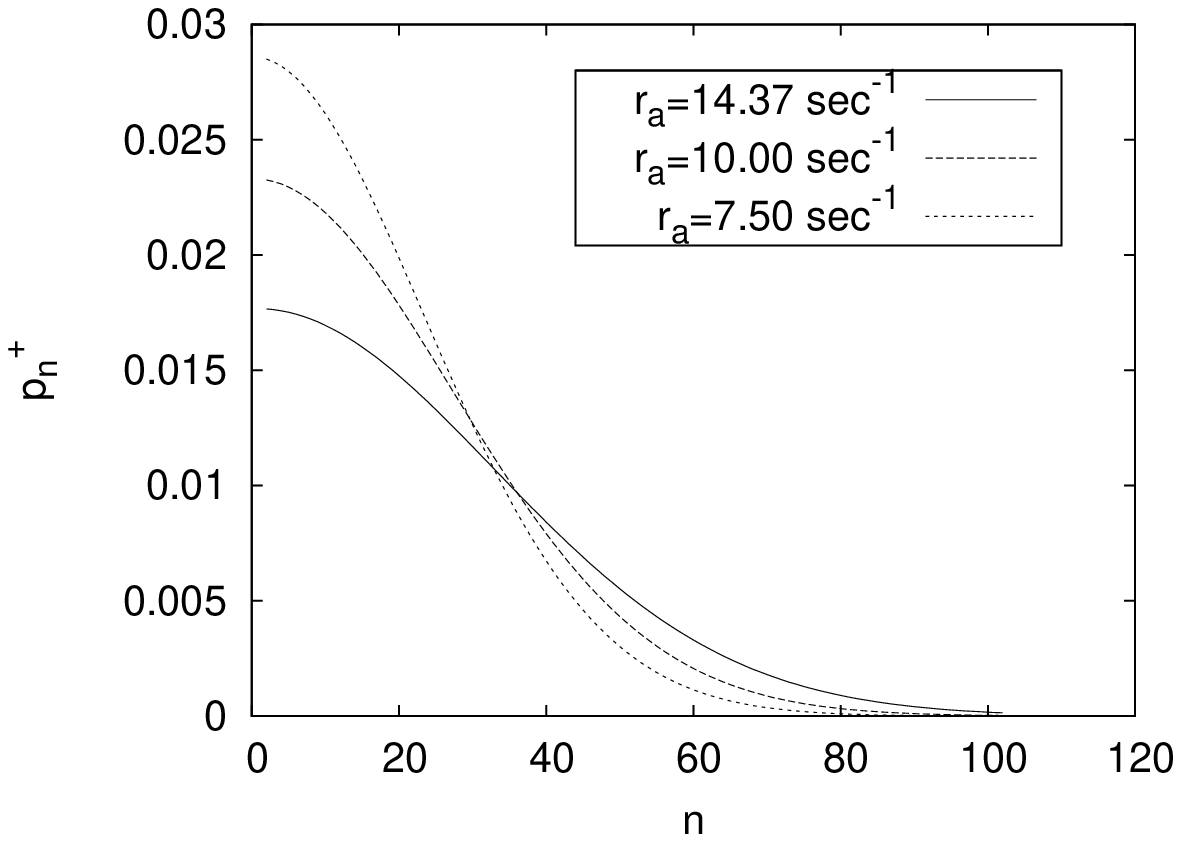}
    \caption{Steady-state  probability 
densities for growing microtubules 
are plotted as a function of length for different 
attachment rates. 
Values of  parameters other than $r_a$ are $r_d=37.5\  s^{-1}$, 
$r_{+-}=0.014 \ s^{-1}$ and $r_{-+}=0.044 \ s^{-1}$.}\label{fig:p1pdiffra}
\includegraphics[height=2.5 in,clip,angle=0]{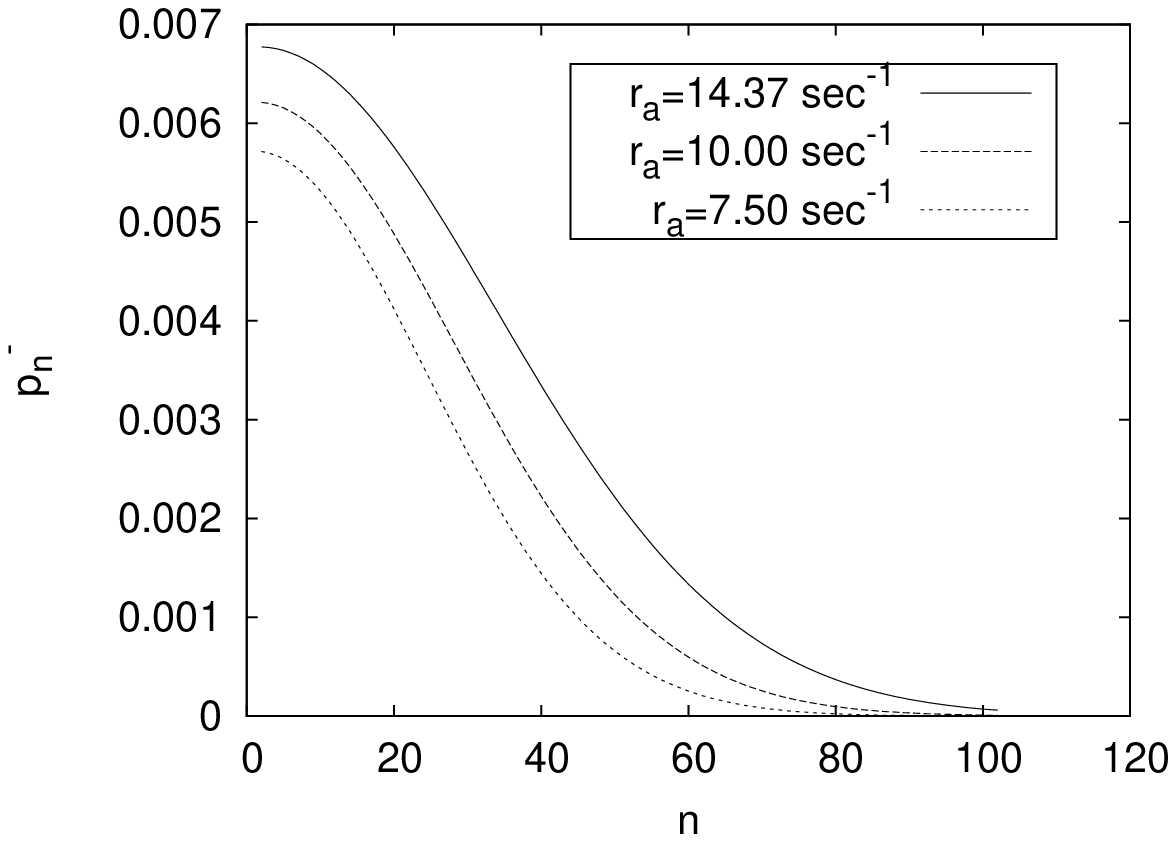}
\caption{Steady-state probability densities for shrinking 
microtubules 
are plotted with length for different attachment rates. 
Parameter values are same as those of  
Fig. \ref{fig:p1pdiffra}.}
\label{fig:p1mdiffra}
  \end{center}
\end{figure}

In order to analyze the evolution equations with length dependent catastrophe 
rate, we consider the  generating functions as given in (\ref{gh})  with 
time dependent probability densities. The generating functions are, in 
this case,   time dependent. 
Summing (\ref{polymer+}) and (\ref{polymer-}) over $n$, we obtain 
the following differential equations for $G(x,t)$ and $H(x,t)$.
\begin{eqnarray}
&&  \frac{dp_1^-(t)}{dt}+x\frac{d H(x,t)}{dt}=
[r_d-(r_d+r_{-+})x] H(x,t)+\nonumber\\
&& 2 r_{+-} x G(x,t)+
 r_{+-} x^2  G'(x,t)+\nonumber\\
&&  (r_{+-} p_1^+-(r_d+r_{-+}) p_1^-),\ {\rm and} \label{diffH1}
\end{eqnarray}
\begin{eqnarray}
&&\frac{d G(x,t)}{dt}=r_a p_1^++(-r_a-2r_{+-}+r_a x)G(x,t)-\nonumber\\
 && r_{+-} x G'(x,t)+r_{-+} H(x,t). \label{diffG2}
\end{eqnarray}
Here, prime denotes derivative with respect to $x$. 
One could also directly consider the steady-state equations and obtain the 
probability densities by a transfer matrix approach. However, since 
 its use is difficult in the length-dependent case, we follow the 
generating function method here. 
Using  Eq. (\ref{polymer-}) for $n=1$,  
all the $x$ independent terms of  equation (\ref{diffH1}) 
can be together expressed in terms of   $p_2^-(t)$.
Taking  Laplace transform of Eqs. (\ref{diffH1}) and 
 (\ref{diffG2}), we have 
\begin{eqnarray}
&&\tilde H(x,s)[s x-r_d(1-h_1 x)]=-r_d \tilde p^-_2(s)+\nonumber \\ 
&& 2 r_{+-} x \tilde G(x,s)+r_{+-} x^2 \tilde G'(x,s), \ {\rm and}\\
&& \tilde G(x,s)[s+r_a (1-x)+2 r_{+-}]+r_{+-} x \tilde G'(x,s)-\nonumber\\
&& r_{-+}\tilde H(x,s)=r_a \tilde p_1^+(s),
\end{eqnarray} 
 where  $h_1=\frac{r_d+r_{-+}}{r_d}$ and 
$\tilde f(x,s)=\int_0^\infty dt\ e^{-st}\ \  f(x,t)$. While obtaining 
these equations, we have assumed $G(x,t=0)=H(x,t=0)=0$ i.e. initially 
there are no growing or shrinking polymers of length $n\ge 2$. 
These two equations can be combined to obtain a
single differential equation for $\tilde G(x,s)$ as
\begin{eqnarray}
&&\tilde G'(x,s)[(r_{+-} s+r_d r_{+-})x^2-r_d r_{+-} x]+
r_d r_{-+} \tilde p_2^-(s)+\nonumber\\
&& \tilde G(x,s)[A+B x+c x^2]=r_a 
[s x-r_d(1-h_1 x)]\tilde p_1^+(s),\nonumber\\ 
 \label{glaplace}
\end{eqnarray}
where 
\begin{eqnarray}
&& A=-s r_d-r_a r_d-2 r_{+-} r_d,\\
&& B=s^2+s r_d+s r_{-+} +r_a s+r_a r_d+ r_a r_d h_1+\nonumber\\
&& 2 r_{+-} s+2 r_{+-} r_d, \ \ {\rm and}\\
&&  C=-r_a s-r_a r_d h_1. \label{abc}
\end{eqnarray}
In order to obtain the steady state behavior, we use the 
final value theorem  which is based on the assumption 
that the steady-state 
generating function $G(x)=G(x,t\rightarrow \infty)$  exists and 
is given by $G(x,t\rightarrow \infty)=\lim_{s \to 0} s \tilde G(x,s)$. 
The same is assumed to be valid for other functions  $\tilde p_1^+(s)$ 
and $\tilde p_2^-(s)$. As $s\rightarrow 0$, Eq. (\ref{glaplace})  becomes
 singular at $x=1$. However, as in the length independent 
case, this singularity 
is a removable one  when the appropriate steady-state expressions 
for $p_1^+$ and $p_1^-$ are used. After these substitutions, the 
steady-state generating function satisfies  the following equation 
\begin{eqnarray}
&& (r_{+-} r_d) x \frac{dG}{dx}+(d_1-d_2 x) G(x)=d_3,\ \  {\rm where }\\
&& d_1=r_d(r_a+2 r_{+-}),\ \ \ d_2=r_a(r_d+r_{-+})\ \  
{\rm and} \ \nonumber\\ 
&& d_3=p_1^+ r_a(r_d+r_{-+}). \label{genfinal}  
\end{eqnarray} 
This inhomogeneous equation has a solution of the form 
$G(x)=G_c(x)+G_p(x)$
 having a complementary part 
\begin{eqnarray}
G_c(x)= C_0 x^{-\frac{d_1}{r_{+-}r_d}} 
\exp[\frac{d_2}{r_{+-}r_d} x] \label{complementaryG}
\end{eqnarray}
with $C_0$ as a constant and a particular integral 
for which 
a series of the form in Eq.  (\ref{gh})  can 
be considered.
Eq. (\ref{genfinal}) has a singularity at $x=0$ and this results in a 
singularity of the complementary solution   at  $x=0$.   
Since a catastrophe rate proportional to the microtubule length 
is expected to  decrease the probability densities than those of 
the length independent case
and a diverging solution for the generating 
function  is unlikely, we choose $C_0=0$. 
The particular solution gives the probability densities as
\begin{eqnarray}
&& p_2^+ =d_3/d_1, \\
&& p_n^+=\frac{d_3}{d_1} \frac{d_2^{n-2}}
{\prod_{m=1}^{n-2}(d_1+m r_{+-} r_d)} \ \ {\rm for}\ \  n>2, 
\label{pnpfinalform}\\
&& p_{n+2}^-= \frac{r_a}{r_d} p_{n+1}^+\ \ {\rm for }\ \ n\ge 0 
\label{pnmfinalform}.
\end{eqnarray}    
Eq. (\ref{pnpfinalform}), can be rewritten in the form 
\begin{eqnarray}
 p_n^+=\frac{d_3}{d_1} \left(\frac{d_2}{(r_{+-} r_d)}\right)^{n-2} 
\frac{\Gamma(d_1/(r_{+-}r_d)+1)}{\Gamma((n-1)+d_1/(r_{+-} r_d))},
\end{eqnarray}
where $\Gamma(x)$ is the usual $\Gamma$ function of argument $x$. 
Hence, for large $n$, the exponential decay of (\ref{pnpexpo}) 
is modified  as  
\begin{eqnarray}
p_n^+ \sim \exp[-\alpha' n-n\ln n], \label{nlogn}
\end{eqnarray}
with $\alpha'=\ln\frac{(r_{+-}r_d)}{d_2}-1$.
As Figs. \ref{fig:valuep1p}  and \ref{fig:valuep1m} show, 
the decay of   $p_n^+$ of $p_n^-$,  dominated by an $\exp[-n\ln n]$ type 
term, is much faster than that of the length independent 
dynamics \cite{numbers}. 
 The ratio $p_n^-/p_{n-1}^+$, however,  remains same as that of 
the length independent case. 
 In Figs. \ref{fig:p1pdiffra} and 
\ref{fig:p1mdiffra}, we have plotted the probability densities 
for different attachment rates. Fig. \ref{fig:p1mdiffra} shows 
an overall increase of  
$p_n^-$ with the attachment rate. This   seems to be a 
consequence of rapid polymerization and hence 
an enhanced transition of microtubules from growing to the shrinking 
phase. 
With the present form of  distributions of growing and shrinking 
microtubules, the average length of  microtubules is given by 
$n\approx 26.4$ 
for $r_a=14.37\ {\rm sec}^{-1}$ and 
other parameter values same as provided in Fig. 
\ref{fig:p1pdiffra}.

In a continuum description of the  microtubule length, one may have 
the following steady-state differential equations \cite{dogterom3} 
\begin{eqnarray}
&& -r_a a \frac{\partial p^+(L)}{\partial L}+
r_{-+}p^-(L)-r_{+-}\frac{L}{a} p^+(L)=0 \label{cont1}\\
&& r_d a\frac{\partial p^-(L)}{\partial L} +r_{+-}\frac{L}{a} p^+(L)-
r_{-+}p^-(L)=0,\label{cont2}
\end{eqnarray}
where $a$ is the length of the  tubulin dimer. 
In case of bounded solutions for $p^+(L)$ and $p^-(L)$, it is found 
that $\frac{p^-(L)}{p^+(L)}=\frac{r_a}{r_d}$ \cite{dogterom3}. 
Eq.   (\ref{pnmfinalform}) and also  the corresponding eq. 
in the length independent case ( Eq. (\ref{ratioindependent}) )
are different from this. 
Although (\ref{pnmfinalform}) and its  continuum analogue
mentioned above   are simple 
balance equations in terms of the rate parameters,
 the subtle difference between them 
becomes important here  unlike the length independent case. 
In order to find the reason for such difference, 
 we do a Taylor expansion of   Eq. (\ref{pnmfinalform}) in  
the dimer length, $a$.
Retaining terms up to first order in $a$,
the ratio  appears as 
$\frac{p^-(L)}{p^+(L)}=\frac{r_a}{r_d} 
(1-a\frac{1}{p^+(L)} \frac{\partial p^+(L)}{\partial L})$. 
Replacing the first order term using Eq.
 (\ref{cont1}), it can be seen that the ratio  
becomes  same as that of \cite{dogterom3} 
provided $r_{+-}/r_a<<a/L=1/n$.
Rather than referring this as a condition for validity of the 
 continuum approximation, it is more appropriate to mention that  this  
condition expresses a competition between the catastrophe and 
the growth rate which, as we shall see below, distinguish two regimes 
with two different kinds of probability distributions.

That the distribution in (\ref{pnpfinalform}) finally becomes 
a Gaussian distribution for $r_{+-}/r_a<<a/L$, can be seen by 
approximating the logarithm of the product in the denominator of 
(\ref{pnpfinalform}) in the following way.
\begin{eqnarray}
&&\ln \prod_{m=1}^{n-2} (d_1+m r_{+-} r_d)=\nonumber\\
&& (n-2) \ln d_1+\sum_{m=1}^{n-2} \ln(1+m \frac{ r_{+-} r_d}{d_1})\approx 
\nonumber\\
&&(n-2) \ln d_1+\int_0^L \ dx\ 
\ln(1+\frac{x}{a} \frac{r_{+-} r_d}{d_1})\approx\nonumber\\
&& (n-2) \ln d_1+\frac{r_{+-}}{(r_a+2 r_{+-})a}\int_0^L dx \ x.\label{gauss}
\end{eqnarray}
Since $r_{+-}/r_a<<a/L$, we have used $\ln(1+x) \approx x$ to 
approximate the integrand. It is straightforward 
now to see that (\ref{gauss}) finally leads to a Gaussian 
distribution for the length.
$\exp[-n \ln n]$ form of Eq. (\ref{nlogn}), on the other hand,
 follows in the opposite situation where $1/n=a/L<<r_{+-}/r_a$. 
 This analysis shows 
that the higher is the growth rate the quicker is the fall of the 
probability distribution of growing polymers.
 This seems to be a  consequence of the fact that 
a rapid growth induces more frequent catastrophe.

\section{Summary}
 It has been found that the catastrophe frequency becomes 
dependent on the microtubule length  due to the activity of 
depolymerases belonging  kinesin-8 family.
Here, we consider a two-state model of growing and shrinking microtubules 
having a catastrophe rate that increases linearly with the length of the 
microtubule. Following a discrete approach, we  study  
how the steady state probability densities $p_n^+$ 
and $p_n^-$ of growing and shrinking polymers, respectively,  of length $n$,
vary with length.  Our results show that, for large $n$, the decay of the 
probability densities
  with $n$ is dominated by an $\exp[-n \ln n]$ type of term
as compared to an  $\exp[-\alpha n]$ kind of decay in the length independent 
catastrophe case.

{\bf Acknowledgement}
Financial support from the   Department of Science 
and Technology,  India   is   gratefully acknowledged.  


\begin{thebibliography}{99}
\bibitem{numbers} Values of the parameters, used for the figures, 
are obtained from  references of experimental works 
provided in \cite{dogterom3}.
\end{thebibliography}
\end{document}